\documentclass[conference]{IEEEtran}
\IEEEoverridecommandlockouts
\usepackage{cite}
\usepackage{amsmath,amssymb,amsfonts}
\usepackage{algorithmic}
\usepackage{graphicx}
\usepackage{textcomp}
\usepackage{xcolor}
\def\BibTeX{{\rm B\kern-.05em{\sc i\kern-.025em b}\kern-.08em
    T\kern-.1667em\lower.7ex\hbox{E}\kern-.125emX}}

\begin{document}

\title{ECU Identification using Neural Network Classification and Hyperparameter Tuning}

\author{\IEEEauthorblockN{Kunaal Verma}
\IEEEauthorblockA{\textit{CECS Department} \\
\textit{University of Michigan}\\
Dearborn, USA \\
vermakun@umich.edu}
\and
\IEEEauthorblockN{Mansi Girdhar}
\IEEEauthorblockA{\textit{CECS Department} \\
\textit{University of Michigan}\\
Dearborn, USA \\
gmansi@umich.edu}
\and
\IEEEauthorblockN{Azeem Hafeez}
\IEEEauthorblockA{\textit{CECS Department} \\
\textit{University of Michigan}\\
Dearborn, USA \\
azeemh@umich.edu}
\and
\IEEEauthorblockN{ Selim S. Awad}
\IEEEauthorblockA{\textit{CECS Department} \\
\textit{University of Michigan}\\
Dearborn, USA \\
sawad@umich.edu}
}

\maketitle

\begin{abstract}

Intrusion detection for Controller Area Network (CAN) protocol requires modern methods in order to compete with other electrical architectures. Fingerprint Intrusion Detection Systems (IDS) provide a promising new approach to solve this problem. By characterizing network traffic from known ECUs, hazardous messages can be discriminated. In this article, a modified version of Fingerprint IDS is employed utilizing both step response and spectral characterization of network traffic via neural network training. With the addition of feature set reduction and hyperparameter tuning, this method accomplishes a 99.4\% detection rate of trusted ECU traffic.

\end{abstract}


\section{Introduction}

Automotive electrical systems consist of several physically isolated ECUs that control various functions and operations of a vehicle, e.g., engine control, traction control. Common in-vehicle communication bus networks include CAN, Local Interconnect Network (LIN), and FlexRay \cite{9257265}. CAN is the most commonly used bus system in automotive networks, and it has proven to be a reliable architecture for the exponential growth of vehicle electrical systems. 

As in-vehicle electrical architectures evolve in both complexity and capability, so does their potential for cybersecurity attacks and intrusion. CAN, unlike more modern architectures, e.g., automotive Ethernet, does not have the inherent capability to handle these new threats to security and safety \cite{8658720}. However, due to its ongoing success and broad institutional knowledge manufacturers will continue to use CAN as the backbone of in-vehicle architecture for the foreseeable future.

In order to maintain CAN's relevancy in the face of newer and more secure electrical architectures, it's security capabilities must be enhanced by novel methods for intrusion detection and mitigation. This becomes even more pressing with the onset of advanced connected-vehicle features such as vehicle sensor networks, autonomous control, and mobile connectivity. In addition, various interactions and interdependencies between several cyber-physical components or ECUs may cause eavesdropping, spoofing or Denial of service (DoS) attacks, hence compromising the system\cite{9583592}. There are also growing concerns of increased remote intrusion with the proliferation of mobile network integration into modern vehicles \cite{bworld2}.


CAN, a vehicle network communication standard using CSMA/CD \cite{CSMA}, allows network messages to be broadcast to every ECU on the network on an open bus architecture. This topology is beneficial for ease of information access across the network, allowing an arbitrary ECU to extract all the information they need from other ECUs while maintaining low wiring cost and design complexity \cite{6025574}. Unfortunately, this also means that intrusive actors with physical access to the vehicle can spoof ECU messages easily.

\begin{figure}[htb]
\centering
\includegraphics[width=3.1in]{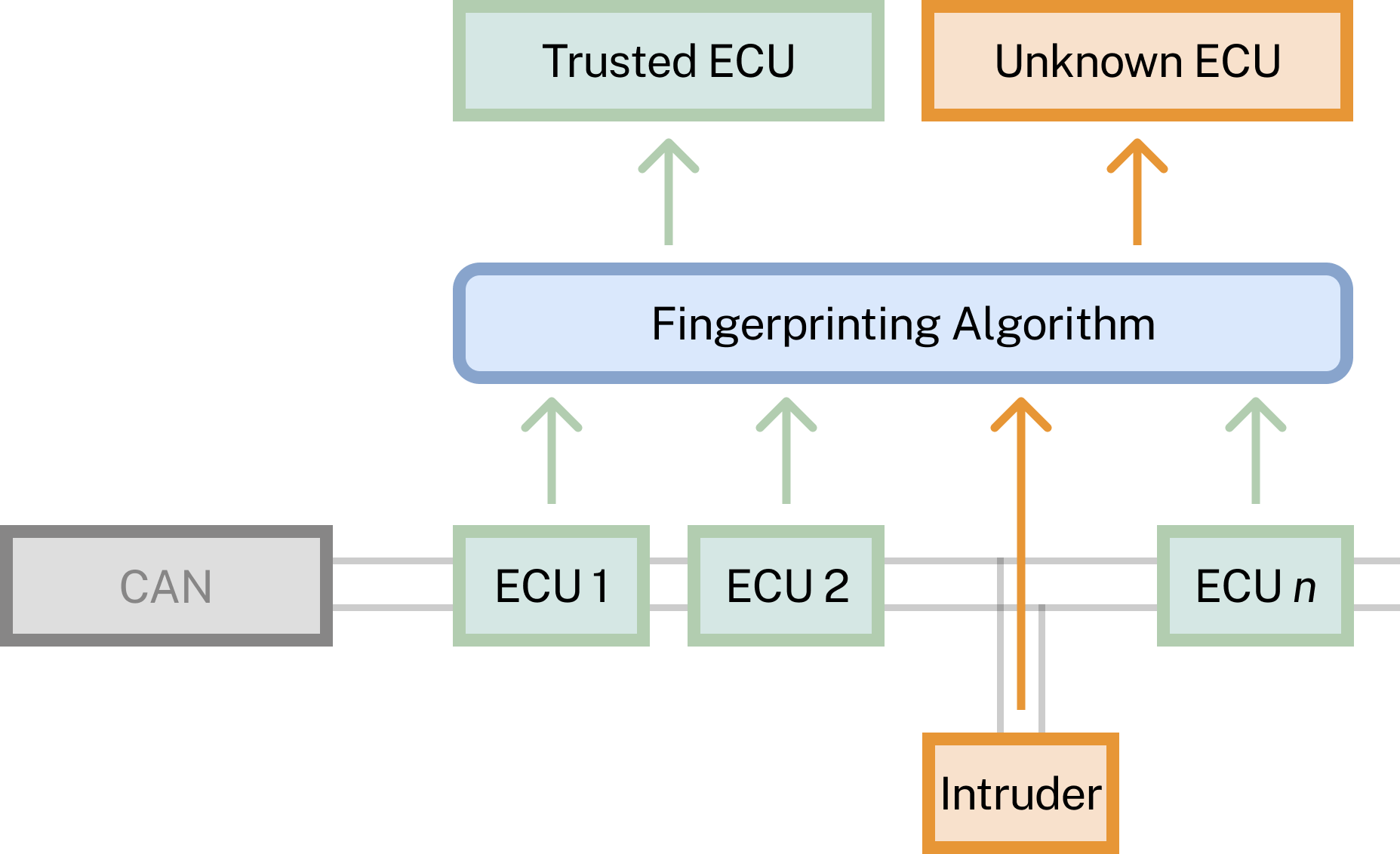}
\caption{$CAN  \; Topology  \; with  \; ECU \;  Fingerprinting$}
\label{fig:NetworkTopology}
\end{figure}

One solution to this problem is by uniquely identifying known ECU messages, and recognizing when unknown ECU messages are present on the network. This can be accomplished in a number of ways, but it is important that the technique used for ECU identification creates enough contrast and uncorrelated behavior with other ECUs so that messages on the network can be classified accurately and with little confusion. Once known and trusted ECUs have been successfully identified and integrated into a detection model, fingerprinting can be employed to detect and isolate unknown ECU signatures, as illustrated in Fig \ref{fig:NetworkTopology}.

One such method, proposed by \cite{hafeez2019}, shows a promising way of fingerprinting network messages. Transient and step-response features are extracted from sampled ECU communications. This feature set is then used to train an ANN to learn which messages are sent by a given ECU, and then identify whether future messages belong to trusted ECUs or adversarial ECUs.

In the proposed work, we investigate further improvements of this method by increasing the feature space used to train the ANN. In addition to signal transient and steady-state characteristics, spectral analysis characteristics are included to further bolster the classification accuracy. In addition, a feature space reduction is performed to optimize performance and remove features that provide neutral or negative influence during the classification of messages.

With this new work we hope to replicate the previous results of \cite{hafeez2019} and improve upon accuracy with our proposed modifications.

\color{black}
\section{Message Detection Methods} \label{sec:MDM}


There are two main categories of classification that are discussed in the current literature on ECU authentication and identification: (i) message authentication code (MAC) \cite{9027298}, and (ii) IDS \cite{9216166}.

MAC-based methods achieve security by encrypting the message payloads of a CAN packet prior to transmission. This family of authentication protocols can lead to high transmission costs, and part cost to each ECU on the network \cite{gierlichs2016cryptographic, hazem2012lcap, doan2017can, ueda2015security, sugashima2016approaches}.

Alternatively, IDS relies on the uniqueness of each ECU's electrical operation and message transmission channel to create unique profiles from messages transmitted by the ECU. This requires a supervisory monitoring ECU or additional compute on a centralized ECU, but the data rate requirements of IDS are much lower than that of MAC, meaning that network bandwidth is not affected by the authentication system. Additionally, there are four sub-classes of IDS that span the current literature:

\textbf{Parameter Monitoring} IDS involves using periodic frequency \cite{taylor2015frequency} and time-of-flight (TOF) (remote frame) characteristics \cite{lee2017otids} of messages to establish a baseline of behavior. Later, new messages are measured against the expected message parameters in order to identify suspicious or fraudulent messages. 

\textbf{Information Theory} IDS relies upon an ordered messaging schedule in which a specific sequence of messages is established. If the sequence is violated, this indicates the presence of an attack. This can prove to be a fairly effective technique as the authors in \cite{wang2017hardware} collected 6.673 million CAN packets for different vehicles, and the experimental results show that CAN messages have low entropy of an average of 11.915 bits.

\textbf{ML} IDS approaches rely upon consistent and regularly observed characteristics for each node in the network. In general, regression and clustering algorithms are trained on signals or observed behaviors from each node separately in order to uniquely identify the node. Then, an advisory system with these known signatures confirm that messages propagating in the network are indeed coming from the authorized nodes.
ML-based methods are effective for detecting network attacks on nearly every level of communication, i.e., from physical to application layer. \cite{wu2019survey} Examples of these methods include Long Short Term Memory (LSTM) \cite{taylor2016anomaly}, Ternary Content Addressable Memory (TCAM) \cite{markovitz2017field}, and CAN Frame Deep Learning \cite{kang2016intrusion}.

\textbf{Fingerprint} IDS involves measuring physical characteristics of message packets sent from other ECUs on the network. There are many characteristics that can be extracted from signals and messages observed on the CAN bus, and therefore it is a method that can be implemented in many different ways.

Previous works have employed different techniques, e.g., clock-based intrusion detection (CIDS) \cite{cho2016fingerprinting}, time and frequency domain features \cite{avatefipour2017}. However, the latter method is the primary focus of this paper, which works on optimizing physical signal feature extraction and selection for categorization of CAN traffic.

\section{A Practical Intrusion Detection Module}

A practical realization of an IDS consists of a microcontroller with a frame buffer that isolates several recessive-dominant transitions during an ECU's communication frame. This should allow the data processing algorithm to isolate at least one single-bit pulse for each ECU, for every period of CAN traffic. The frame buffer could then be processed to classify the origin of the message.

\begin{figure}[htb]
\centering
\includegraphics[width=2.6in]{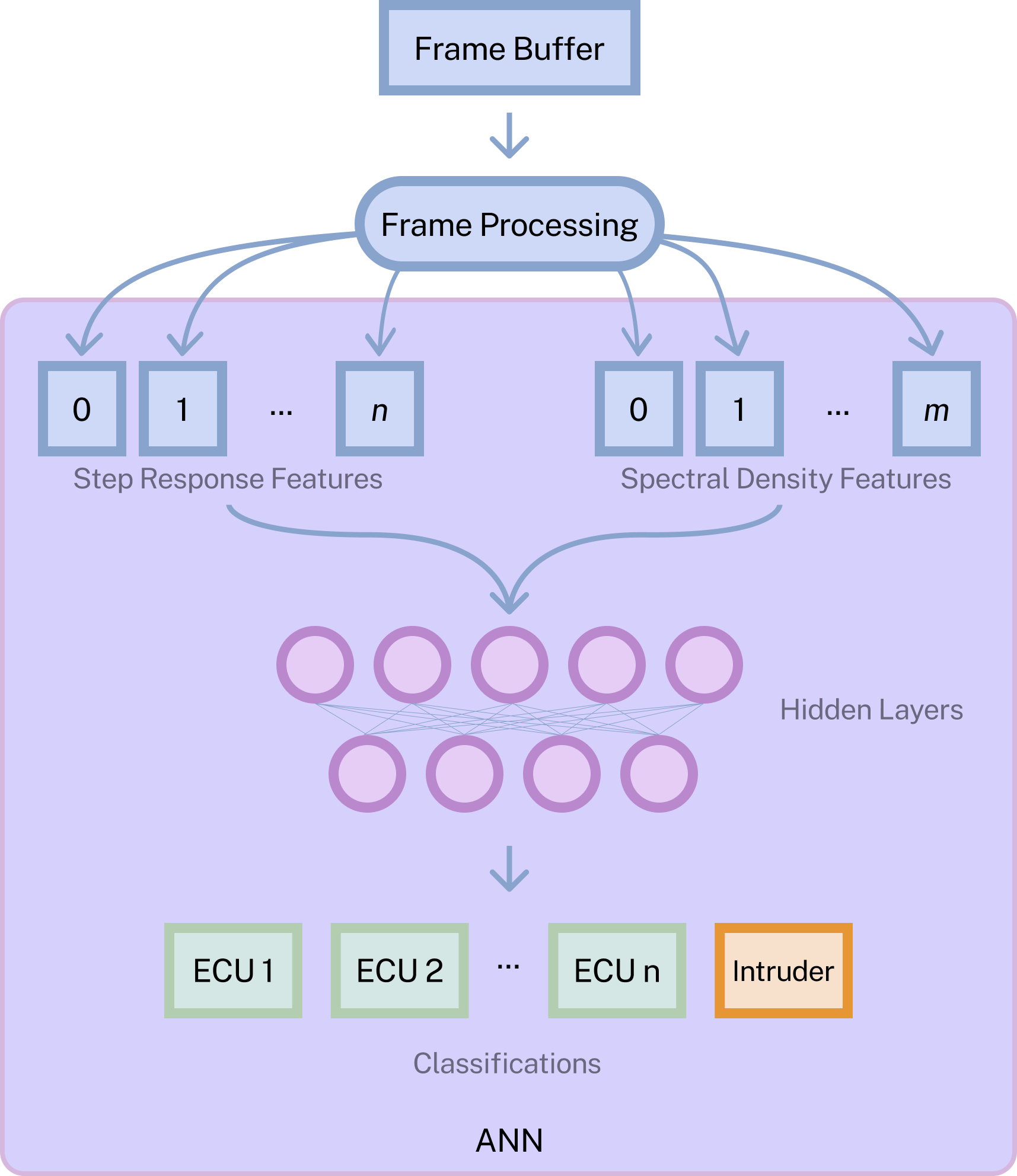}
\caption{$Algorithmic  \; Data \; Flow$}
\label{fig:CANBitLogic}
\end{figure}

Additionally, the microcontroller could be designed to perform reinforcement learning of the neural network in order to adjust to changing physical conditions of the ECU and harness wiring over time, as well as for addressing the dynamic behavior of intruders. A cheaper alternative could utilize other compute resources in the vehicle for network training, and using the IDS microcontroller to hold the neural network in memory.

In this paper, we will discuss a static / off-line data processing method to produce the desired classification of network messages.

\section{Dataset Preparation} \label{sec:Dataset}
The dataset used in this paper originates from the one used in \cite{hafeez2019}, which includes CAN-High (CANH) protocol data from 7 ECUs. Each ECU is recorded 30 times on average, with roughly 600 samples per record at 20 kHz baud rate. The records provided show a regular clock pulse from CANH dominant bit (logic 1) to recessive bit (logic 0), with an 86\% duty-cycle of the dominant bit. CANH has two logical voltage levels, 3.5V for the dominant bit and 2.5V for the recessive bit \cite{9698889}. Analyzing the CANH signal using a variety of fingerprinting feature sets allowed us to create ML classifiers of varying accuracy.

One key part of conditioning our static data required employing \textbf{Out-Of-Bound Pulse} protection. This is a pre-processing step used to isolate clean pulses that in turn produce a better result in algorithm performance.

For a peak index vector $\textbf{i} = [i_1\;i_2 \cdots i_n]\;\forall i \in \mathcal{W}$ and peak times $\textbf{t}(i): (\mathcal{W} \rightarrow \mathcal{R}$) sec., we check to see if any period $t_p = t(i_n) - t(i_{n-1})$ between peaks is less than the expected signal interval $t_p^{(avg)} = \frac{1}{F_s}$. $F_s$ is the sampling frequency. For those indices that do not meet our desired criteria, we remove them leaving only the desired peak index vector $\textbf{i}^*$ and desired peak time vector $\textbf{t}^*$ s.t.
\begin{align*}
    \textbf{t}_p &=
        \begin{bmatrix}
            (t_1 - t_0) & (t_2 - t_1) & \cdots & (t_n - t_{n-1})
        \end{bmatrix}\\
    \textbf{t}_p^* &=
        \begin{bmatrix}
            \textbf{t}_p & t_p^{(avg)}
        \end{bmatrix}\\
    \textbf{t}^* &= \textbf{t} \; | \; \textbf{t}_p(i) \geq t_p^{(avg)} \; \forall i\\
    \textbf{i}^* &= \underset{i} \arg \; \textbf{t}^*(i) > 0 \; \forall i
\end{align*}

\section{Feature Selection and Extraction}


In the previous work \cite{hafeez2019}, the feature set used for classification of specific channels relied upon step response signal characteristics. We repeated this analysis (Method 1 or M1) and used it as our baseline for performance.

In addition to step response characteristics, we chose to also include features generated from Spectral Analysis of the sampled signal pulses. Each pulse can be evaluated through the use of the Discrete Fourier Transform (DFT). With this frequency spectrum representation of our signal we also extracted other unique features that could provide improved performance to our neural network model.

\subsection{Rising and Falling Edge Detection}

An important tool for the analysis of step response characteristics is rising edge detection. For our purposes, we utilized a moving average filter with a window size of 6 samples, adjusted the sample values by a threshold value, and then used zero-crossing detection to determine the time and indices of each peak.

In practice, after removing DC bias from the signal, we found that $\pm 0.2$ V provided accurate peak index output for our dataset.

\begin{figure}[htb]
\centering
\includegraphics[width=3.0in]{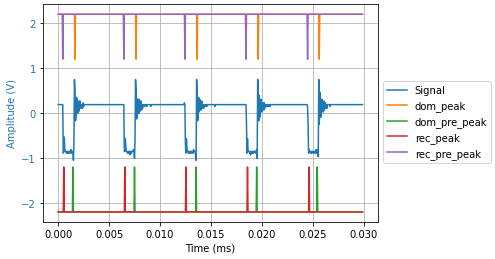}
\caption{$Edge\;Detection\;Output$}
\label{fig:Edge}
\end{figure}

Similarly, we also are interested in the falling edge of each pulse to measure other features from a signal. A similar process is applied, but in addition we also dynamically paired leading and falling edges to isolate dominant and recessive pulses. Fig. \ref{fig:Edge} illustrates both processes on a multiple pulse measurement.
 
\subsection{Peak Time (T$_p$)}
We calculated peak time by utilizing the paired rising and falling edge indices, as well as the sampling time of the signal:
\begin{equation*}
    T_{peak} = (i_{rise} - i_{fall}) \cdot T_s   
\end{equation*}

\subsection{Steady-State Value (SSV)} 
Measurement of the SSV required a moving average filter window size approximately equal to the shorter bit pulse length, followed by threshold detection. In this case, the window size was equivalent to the recessive bit pulse duration of roughly 20 samples.

\subsection{Steady-State Error (SSE)} \label{sec:SSE}
In order to calculate the SSE, we chose the closest ideal voltage using the SSV of the current pulse ($V_{ideal}(i) =$ 2.5V if $i \in \textbf{Rec}$, 3.5V if $i \in \textbf{Dom}$) and found the difference.

\subsection{Percent Overshoot (\%OS)}
Percent overshoot measures the amount of overshoot on the rising edge of a pulse above $SSV$ and is defined as,

\begin{equation*}
    \%OS = 100\% \times \frac{\text{Peak Amplitude} - SSV}{SSV}
\end{equation*}

\subsection{Settling Time (T$_s$)}
\begin{figure}[htb]
\centering
\includegraphics[width=3.2in]{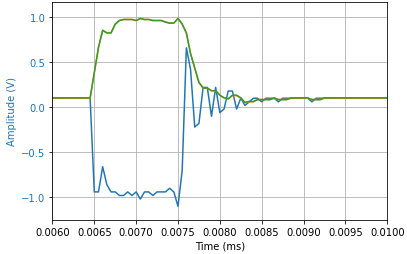}
\caption{$Using\;Signal\;Envelope\;to\;find\;Settling\;Time$}
\label{fig:Settling}
\end{figure}

 Calculating settling time required using threshold detection on a smoothed signal envelope, an example is shown in Fig. \ref{fig:Settling}. A 5\% detection threshold was used to avoid poor measurement consistency caused by aliasing and noise in the signal.

\subsection{Rise Time (T$_r$)}
Rise Time is the amount of time for the rising edge of a pulse to reach $SSV$. Determining this value is similar to the process for finding settling time, using $SSV$ as the threshold value on the rising edge of each pulse.

\subsection{Delay Time (T$_d$)}
Delay Time is the amount of time for the rising edge of a pulse to reach $0.5\,SSV$. Determining this value is similar to the process for finding settling time, using $0.5\,SSV$ as the threshold value.

\begin{figure}[htb]
\centering
\includegraphics[width=3.2in]{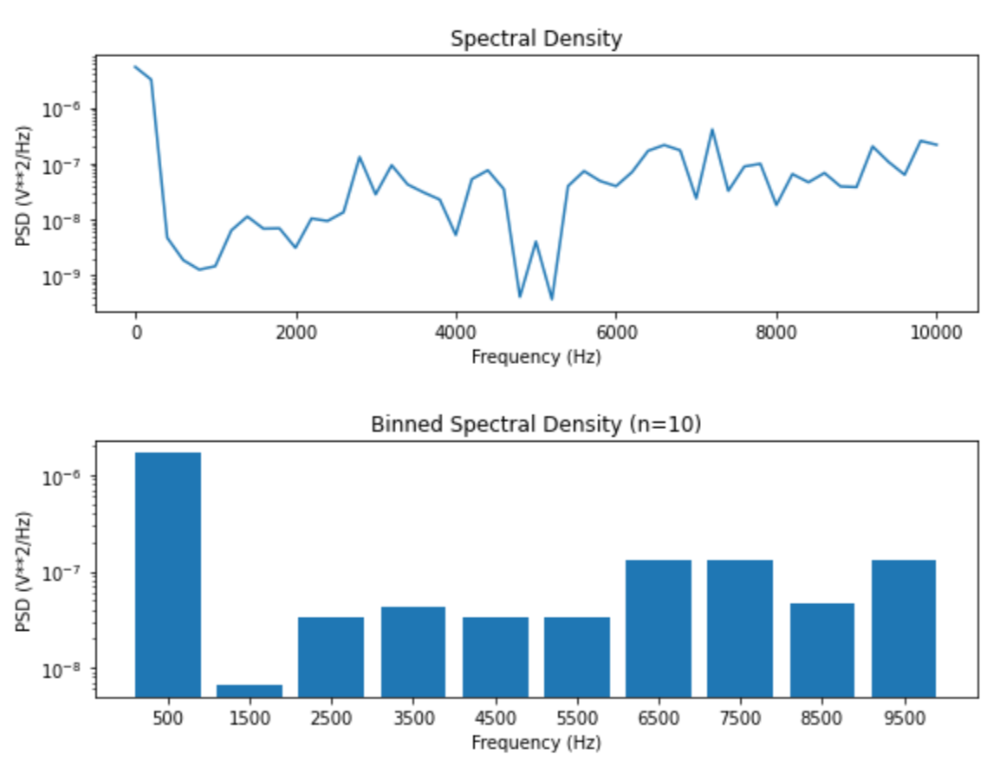}
\caption{$Bin-Reduced\;Power\;Spectral\;Density$}
\label{fig:PSDBin}
\end{figure}

\subsection{Power Spectral Density (PSD or SD)}
The method for frequency spectrum transformation of the signal data used in this work utilizes Welch's method \cite{welch1967} to perform an efficient FFT of a sequential input vector (samples series data in our case). This algorithm is available in the \textit{scikitlearn} library.

We could have utilized every value of the \textit{PSD} magnitude vector (~200 values) in our ML model training, however this could have lead to a high computational cost. Instead, we chose to adopt frequency binning as a means of capturing the spectral shape of the pulse's frequency domain information, as shown in Fig \ref{fig:PSDBin}.


\subsection{Signal-To-Noise Ratio (SNR)} \label{sec:SNR}
\begin{figure}[htb]
\centering
\includegraphics[width=3.2in]{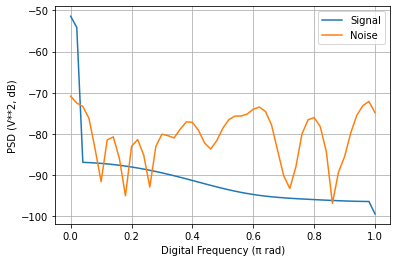}
\caption{$Signal\;and\;Noise\;Power\;Spectral\;Density$}
\label{fig:SNR}
\end{figure}

SNR measures the ratio between the power of the ideal pulse signal (step function) and the power of the signal noise (measured signal minus the ideal step function). It is shown in Fig. \ref{fig:SNR}.

\begin{align*}
    SNR &= 10 \times log_{10} \frac{ \textbf{SUM}(PSD_{signal}) }{ \textbf{SUM}(PSD_{noise}) }
\end{align*}

\subsection{Mean Frequency (MNF)}
Mean frequency can be calculated directly from the \textit{PSD} and frequency vector data produced by the FFT algorithm \cite{angkoon2012} as,

\begin{equation*}
    MNF = \frac{ \sum_{j=1}^M f_j P_j }{ \sum_{j=1}^M P_j }
\end{equation*}

\subsection{Median Frequency (MDF)}
Median frequency is the frequency at which energy is split equivalently in the \textit{PSD} of the pulse \cite{angkoon2012}. The \textit{PSD} is normalized, cumulatively summed, offset by 50\% of the signal energy, and an \textbf{abs} value is applied.  Median frequency is found by locating the frequency at which a minima occurs in this resulting vector.

\begin{equation*}
    \sum_{j=1}^{MDF} P_j = \sum_{j=MDF}^{M} P_j = \frac{1}{2} \sum_{j=1}^{M} P_j
\end{equation*}


\subsection{Feature Extraction}
All the acquisition functions detailed above were processed for each signal recording. The resultant database was then parsed for pertinent feature data and copied to a data dictionary, in addition to file metadata (CAN Id, Channel Length (m), Channel Medium, ECU Record Id, and Filepath). Once a record was created for each file, the data dictionary was used to produce the feature subsets that are fed to our ANN.


\section{ML Methodology}
The dataset was split into a training (70\%) and test set (30\%) for each run. In early development static random number generation seeds were utilized. This helped to maintain a consistent understanding of performance across different subsets of features. After generalized performance was understood, the training and datasets were completely randomized, run to run, in order to remove any training bias and overfitting that could influence performance results.

For all feature sets in this work, we trained a multi-layer perceptron neural network model (MLP Classifier from \textit{scikit-learn}) to predict classifications for the ECUs in our testing set. This model utilizes the Adam solver \cite{kingma2017adam}.

For initial training, the hyperparameter values were chosen to be quite large in order to allow the solver to converge to a maximal solution. Early training typically used approximately 1000 hidden layers and 3000 epochs.

After the model was trained, the remaining test records were used to measure performance of the model. A confusion matrix was constructed with the actual values of the test set, and the predicted values of the model.

\section{Feature Tuning}
Two methods are evaluated initially using the feature extraction data,

\begin{enumerate}
    \item Step Response Features (M1) \cite{hafeez2019}
    \item Step Response + Spectral Analysis Features (M2)
\end{enumerate}

\begin{figure}[htb]
\centering
\includegraphics[width=3.2in]{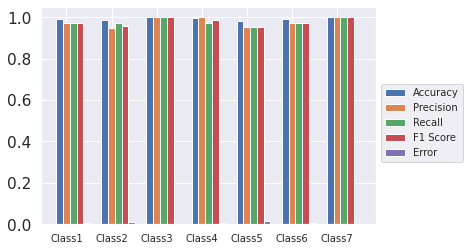}
\caption{$Sample\;M2\;Performance\;Metrics$}
\label{fig:M2pm}
\end{figure}

\begin{figure}[htb]
\centering
\includegraphics[width=2.6in]{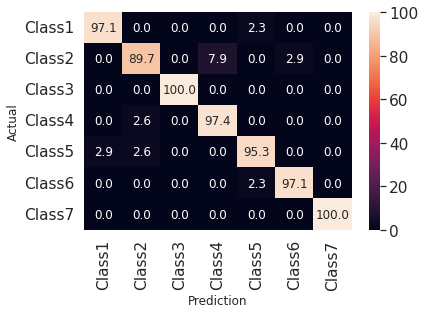}
\caption{$Sample\;M2\;Confusion\;Matrix\;(Percentage\;\%)$}
\label{fig:M2cm}
\end{figure}

Figs. \ref{fig:M2pm} and \ref{fig:M2cm} show sample results for M2. M1 produced an average ECU classification accuracy of 96.85\%, and M2 produced an average accuracy of 98.28\% in the initial trials.

\begin{figure}[htb]
\centering
\includegraphics[width=2.25in]{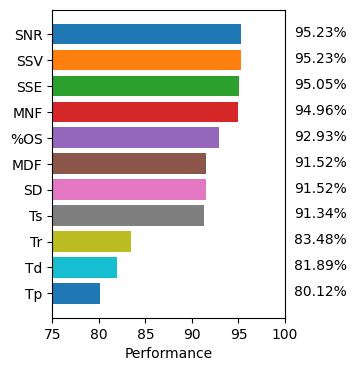}
\caption{$Feature\;Accuracy\;Ranked$}
\label{fig:Rank}
\end{figure}

With M1 performance as a baseline, we looked to improve accuracy by fine tuning the feature set. Neural network training was performed for each individual feature. We then ranked the features by accuracy to identify the best performers (Fig \ref{fig:Rank}).

\begin{figure}[htb]
\centering
\includegraphics[width=2.35in]{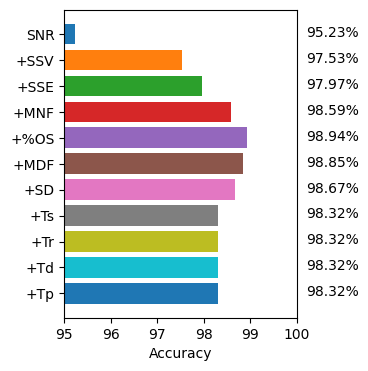}
\caption{$Cumulative\;Feature\;Addition$}
\label{fig:CumRank}
\end{figure}

We then added features one by one to successive neural networks until we observed diminishing returns in accuracy (Fig \ref{fig:CumRank}).

The final feature set includes only $SNR$, $SSV$, $SSE$, $MNF$, and $\%OS$. Overall, we were able to exclude features that did not provide significant value to the overall performance of the neural network model, hence reducing computational cost in the training of the model.

\section{Hyperparameter Tuning}

Although in earlier testing we picked large values for our neural network hyperparameters (maximum epochs of 3000, maximum hidden layers of 1000) we wished to see if there were points of diminishing returns for both.

We found that epoch performance converged at around 500 iterations. Hidden layer performance was variable and could reach optimality from 100-1000 layers. These points/ranges were chosen as the default hyperparameter values and adjusted to maximize accuracy performance for each feature set.

\section{Results}
With our final feature sets chosen, we ran 20 randomized trials using the MLP Classifier, with a hyperparameter optimization of 275 hidden layers and 500 epochs for both M1 and M2. For M1, this led to an average accuracy of 97.17\% (\textbf{M1opt}). A similar optimization was performed with the M2 feature set, resulting in higher performance with a 99.4\% average accuracy (\textbf{M2opt}).

\begin{figure}[htb]
\centering
\includegraphics[width=3.2in]{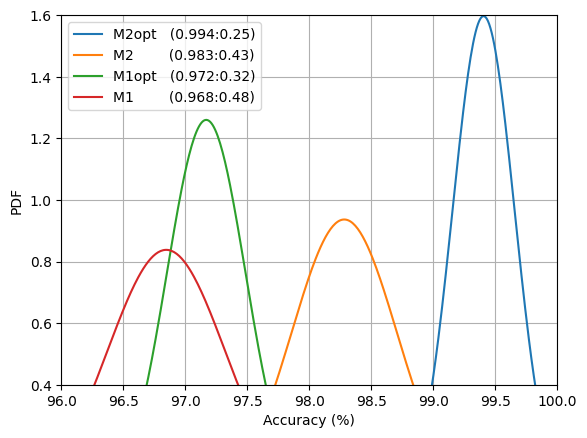}
\caption{$Comparison\;of\;Methods\;(avg:std\_dev)$}
\label{fig:final}
\end{figure}

\section{Conclusion}
\label{sec:concl}


This paper adds new features to Fingerprint-based IDS and optimizes feature selection through a single pass of feature reduction. New features include signal characteristics from the domain of spectral analysis, in particular, PSD, SNR, mean frequency, and median frequency. The feature reduction pass performs a quasi-heuristic search of feature combinations in order to maximize the ratio of performance to computation. Another quasi-heuristic search process is used to optimize hyperparameter values for the multi-layer perceptron neural network. We compare all of the primary methods of this paper in Fig. \ref{fig:final}. The accumulated channel detection accuracy of the best performing method is 99.4\%.


\bibliographystyle{IEEEtran}

\end{document}